\documentstyle[12pt]{article}
\begin{document}

\sf
\title{\bf PET Detectors with 0.4-mm Depth-of-Interaction Resolution}
\author{{\Large Andrei Yu. Semenov} \\ \\ semenov\textunderscore andrei@yahoo.com}
\date{April 9, 2018}
\maketitle
\section{Introduction}

Presice measurements of the photons conversion points in the scintillators are required to achieve a high spatial 
resolution of the PET system. 
I have developed a new method of reconstruction of the depth-of-interaction information for PET detectors with 
dual-side readout. The depth-of-interaction and energy resolutions from Monte-Carlo simulations are presented in 
this paper. The new method allows to reach depth-of-interaction resolution that is about 0.4~mm ($\sigma$) 
 [or about 1.0~mm (FWHM)] for 10-mm long LYSO scintillator. If the precise measurement of the primary photon energy is not a high priority, the new method can be tuned to achieve even better results for the DOI resolution that is better than 0.3~mm ($\sigma$) [or better than 0.7~mm (FWHM)].

At the moment, the developed new method is a trade secret, and it is available for sale. The potential buyers of 
this method should contact the author of this paper via e-mail.

\section{``Standard'' Ratio Method and ``New'' Method}

The proposed method of the measurement of the depth-of-interaction (the ``new'' method) as well as the conventional
``standard'' ratio method (to be compared with the ``new'' method) are based on the comparison of the amplitudes of
the light signals collected from the opposite edges of the LYSO scintillator. 

\begin{figure}[htb]
\vspace*{55mm}
\begin{center}
\includegraphics{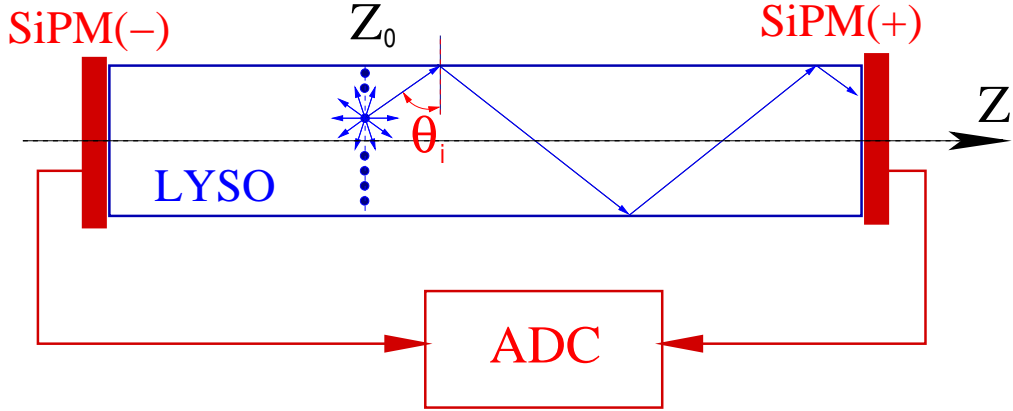}
 \end{center}
\caption{Schematic view of the simulated setup for the ``standard'' ratio method.}
\label{fig:setup}
\end{figure}

A schematic view of the setup for the  ``standard'' ratio method is shown in Fig.~\ref{fig:setup}. The
scintillation light from LYSO crystal is collected on two photodetectors (SiPMs) that are coupled optically on the
opposite sides of the crystal. Detailed parameters of the photon detector can be found in the
Section~\ref{sec:simulation}. 
The signals from the photodetectors are digitized with ADC, and the Z-position of 
the 0.511-MeV photon conversion (depth-of-interaction) is calculated from the ratio of the signal amplitudes.

In comparison with the ``standard'' ratio method, the ``new'' method contains minor changes in the detector as well
as in the signal processing. No change in the readout electronics is required (compared to the ``standard'' ratio
method).

\section{Simulation}
\label{sec:simulation}

The simulation is done for the LYSO scintillator with 1.2$\times$1.2-mm$^2$ cross-section (X-Y)  and lengths (Z)
that varies from 10~mm to 50~mm. I assume the surfaces of the scintillator to be ``perfectly'' polished (viz., no
roughness of the surfaces is introduced in the simulation).  The 1.2$\times$1.2-mm$^2$ output windows are optically
coupled with photodetectors of 50\% photo-detection efficiency (PDE) that corresponds to the characteristics of
J-Series SENSL SiPMs~\cite{pde}.   The optical parameters of the scintillator in the simulation are chosen to be close to the
typical parameters of the LYSO crystals~\cite{refr,att}: $n_1$=1.82 for the refractive index, $\lambda$=15~cm for the "bulk"
attenuation length, and 26,000 photons-per-MeV for the light conversion factor.   

\begin{figure}[htb]
\vspace*{55mm}
\begin{center}
\includegraphics{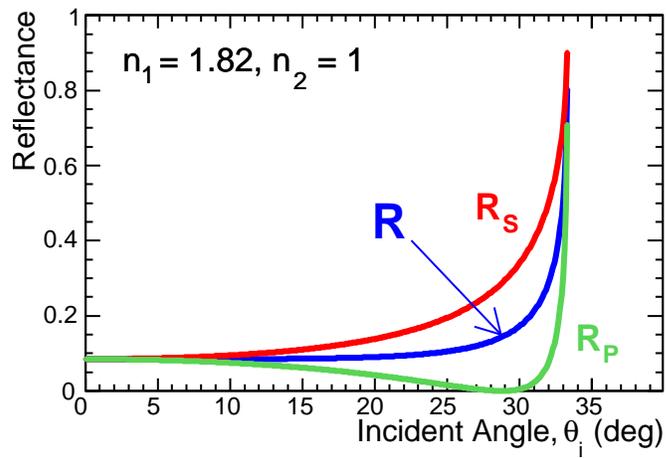}
 \end{center}
\caption{Reflectances for the s- and p-polarized light ($R_S$ and $R_P$) as well as the total reflection probability
$R$ as a function of incident angle $\theta_i$.}
\label{fig:refl}
\end{figure}

The simulation of the spatial and energy resolutions is done as a function of Z-coordinate. For each fixed $Z_0$
position, 5000 conversion points of 0.511-MeV photons (``events'') are uniformly distributed in the
1.2$\times$1.2-mm$^2$ $X-Y$ cross-section of the LYSO scintillator (see Fig.~\ref{fig:setup}).   The simulation of each ``event'' starts with
a seed of $N$  ``optical'' photons, where $N$ is a random number from the Poisson  distribution with the mean value
of 13286 (=26000~photons/MeV $\times$ 0.511~MeV); these photons are originated in the 0.511-MeV photon conversion
point and have  uniformly-distributed directions. After that, I trace each of the ``optical'' photons through the
scintillation  material.
If the photon meets the scintillator wall, I assume the total internal reflections for the incident
angles $\theta_{i}$ that are bigger than the ``critical angle'' $\theta_{cr}$ ($=arcsin(1/n_1)$). 
For the photons with the  incident angles that are less that the ``critical angle'', I assume the reflection with
the probability:
\begin{equation}
R = 0.5 \cdot (R_S + R_P), \label{eq:1}
\end{equation} 
where $R_S$ and $R_P$ are the reflectances for the s- and p-polarized light that are expressed through
Fresnel equations:
\begin{equation}
R_S = \left[\frac{n_1 \cdot cos(\theta_i) - \sqrt{1 - (n_1 \cdot sin(\theta_i))^2}}
{n_1 \cdot cos(\theta_i) + \sqrt{1 - (n_1 \cdot sin(\theta_i))^2}}\right]^2 \label{eq:2}
\end{equation}
\begin{equation}
R_P = \left[\frac{n_1 \cdot \sqrt{1 - (n_1 \cdot sin(\theta_i))^2} - cos(\theta_i)}
{n_1 \cdot \sqrt{1 - (n_1 \cdot sin(\theta_i))^2} + cos(\theta_i)}\right]^2 \label{eq:3}
\end{equation}
The reflection probability $R$ as well as the reflectances $R_S$ and $R_P$ as functions of the $\theta_i$ are shown in
Fig.~\ref{fig:refl}.
One should understand that the Formula~\ref{eq:1} is not exact for the optical photons after one or more
reflections because the photon isn't unpolarized anymore; nevertheless, the probability for the photon to be
reflected after the meeting the scintillator wall (at $\theta_i < \theta_{cr}$) is less than 10\% that makes the
multiple-reflection probability very small.

For the photons that were traced to the output windows, I applied "survival/conversion" probability $Prob$ that
reflects the attenuation in the scintillator material and conversion of photons to photoelectrons (or avalanches
for SiPMs):
\begin{equation}
Prob = PDE \cdot exp(-L/\lambda), \label{eq:4}
\end{equation}
where $L$ is the total path that the optical photon travels to reach the output window.

The number of ``registered'' photons (or avalanches in SiPM) in the left~(-) or right~(+) SiPM are proportional to the
amplitude of the signals in the correspondent SiPMs.

\section{Calibration Procedures and the ``Standard'' Ratio \\Method Results}

\begin{figure}[htb]
\vspace*{55mm}
\begin{center}
\includegraphics{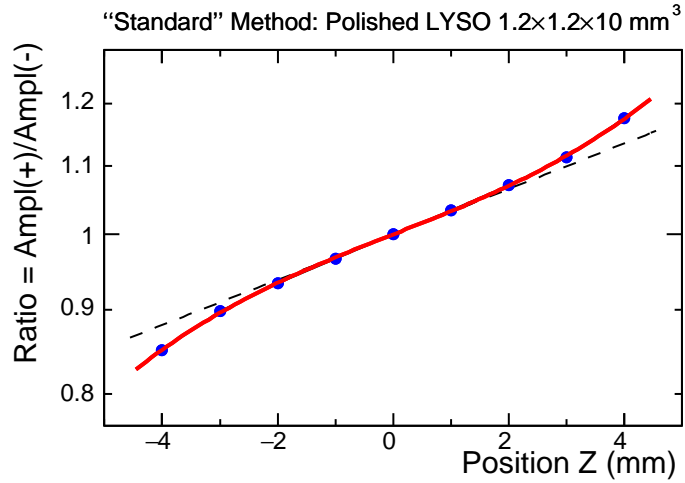}
 \end{center}
\caption{Average ratio of amplitudes from SiPMs as a function of 0.511-MeV photon conversion Z-position (for
the ``standard'' ratio method). The Y-axis has a logarithmic scale.}
\label{fig:ratio}
\end{figure}

An average ratio of the amplitudes of signals from the right and the left SiPMs as a function of 0.511-MeV photon
conversion Z-position (for the ``standard'' ratio method) is shown in Fig.~\ref{fig:ratio}. Near the edges of
scintillator, the ratio deviates from the single-exponent dependence that is shown as a dashed line. To provide accurate
reconstruction of the depth-of-interaction, I fit the simulated ratio data in Fig.~\ref{fig:ratio} with the
formula (shown in Fig.~\ref{fig:ratio} as a solid red line):
\begin{equation}
R(Z) = exp((Z + k \cdot Z^3) / \lambda) ,
\end{equation}
where $k$ and $\lambda$ are the calibration parameters from the fit: $\lambda$ represents the half of the effective
attenuation length in the scintillator, and the parameter $k$ is responsible for the deviation of the signal amplitude
ratio from the single-exponent decay on the edges of the scintillator. Here I use the same simulation data for the
calibration; to calibrate the real-life PET detector, the special calibration data taken with the collimated
$\gamma$-source at few known Z-positions will be needed. 

\begin{figure}[htb]
\vspace*{58mm}
\begin{center}
\includegraphics{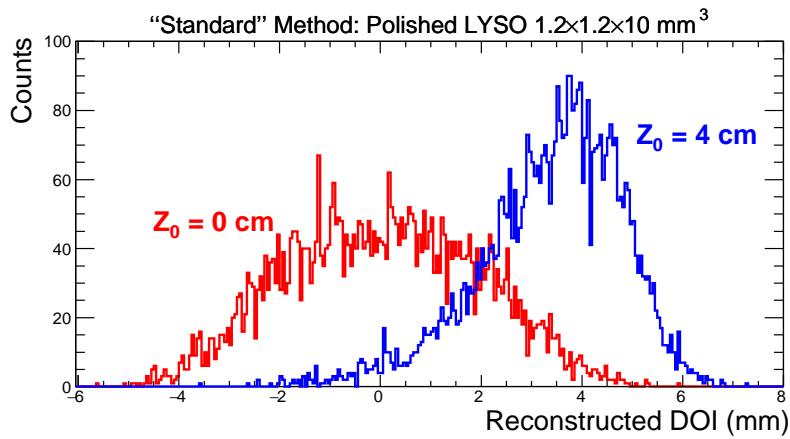}
 \end{center}
\caption{Depth-of-Interaction distributions that are reconstructed with the ``standard'' ratio method for
1.2$\times$1.2$\times$10~mm$^3$ LYSO scintillator at $Z_0$ = 0 and 4~mm. }
\label{fig:standarddoi}
\end{figure}

After the $k$ and $\lambda$ parameters are established, I use them for event-to-event reconstruction of the
depth-of-interaction (DOI) of the primary photon in the scintillator from the ratio $Amp(+)/Amp(-)$ of the amplitude of signals from
the right and the left SiPMs: 
\begin{equation}
DOI = (A + C)^{1/3} + (A - C)^{1/3} ,
\end{equation}
where
\begin{equation}
A = -0.5 \cdot \lambda \cdot ln[Amp(+)/Amp(-)] / k
\end{equation}
and
\begin{equation}
C = \sqrt{A^2 + 1/(27 k^3)}
\end{equation}
DOI distributions reconstructed with the ``standard'' ratio method for 1.2$\times$1.2$\times$10~mm$^3$ LYSO
scintillator at $Z_0$ = 0 (center of the scintillator) and 4~mm (edge of the scintillator) are shown in
Fig.~\ref{fig:standarddoi}. Please note that the distribution at the edge of the scintillator has more compact core
(has smaller FWHM or $\sigma$ resolution) compared to the distribution at the scintillation center, but it has a long
asymmetric "tale". I found that that specific shape of the DOI distribution at the LYSO edge is the result of the
Fresnel reflections; if I ``switch the Fresnel reflections off'' in the simulation (viz., artificially assume $R=0$
in the Eq.~\ref{eq:1}), the DOI distribution at the LYSO edge becomes very similar to the one at the center.
Another observation here is that the reconstructed DOI distributions are peaked at the ``correct'' Z-positions, but
significant distribution widths do not allow precise DOI reconstruction on event-by-event basis with ``standard''
ratio method.

\begin{figure}[htb]
\vspace*{55mm}
\begin{center}
\includegraphics{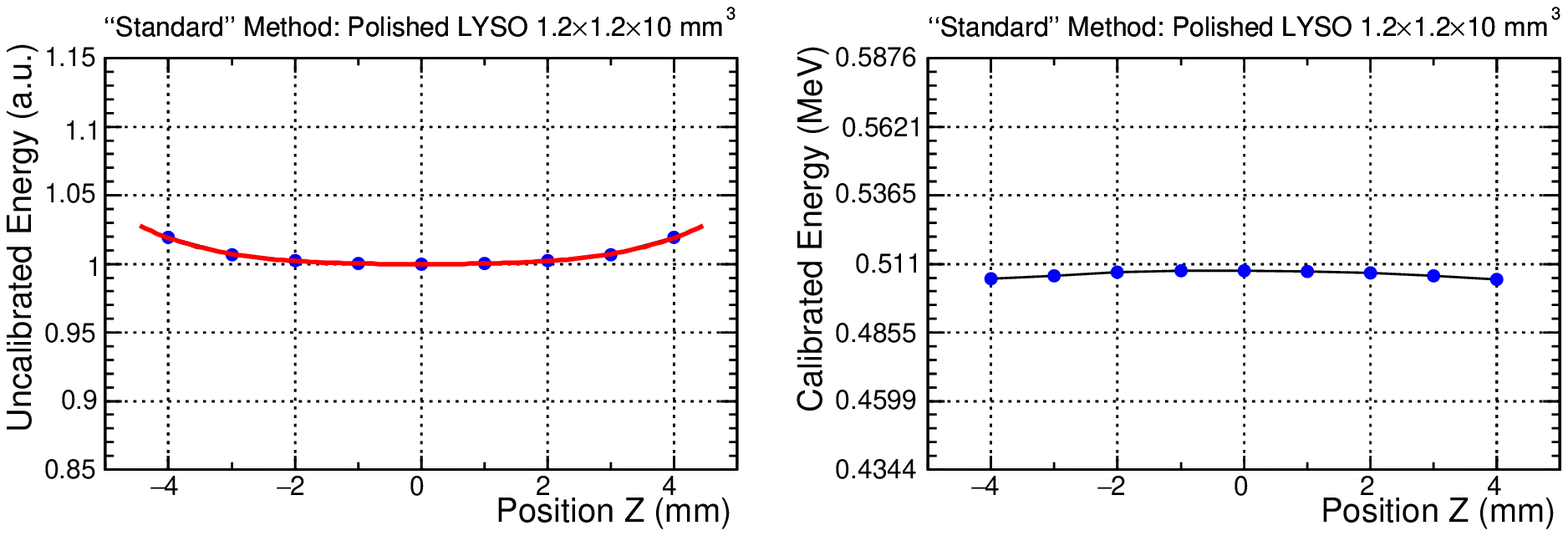}
 \end{center}
\caption{Average energy reconstructed with the ``standard'' ratio method before (left panel) and after calibration (right
panel) as a function of 0.511-MeV photon conversion position.}
\label{fig:standardenergyZ}
\end{figure}

\begin{figure}[htb]
\vspace*{55mm}
\begin{center}
\includegraphics{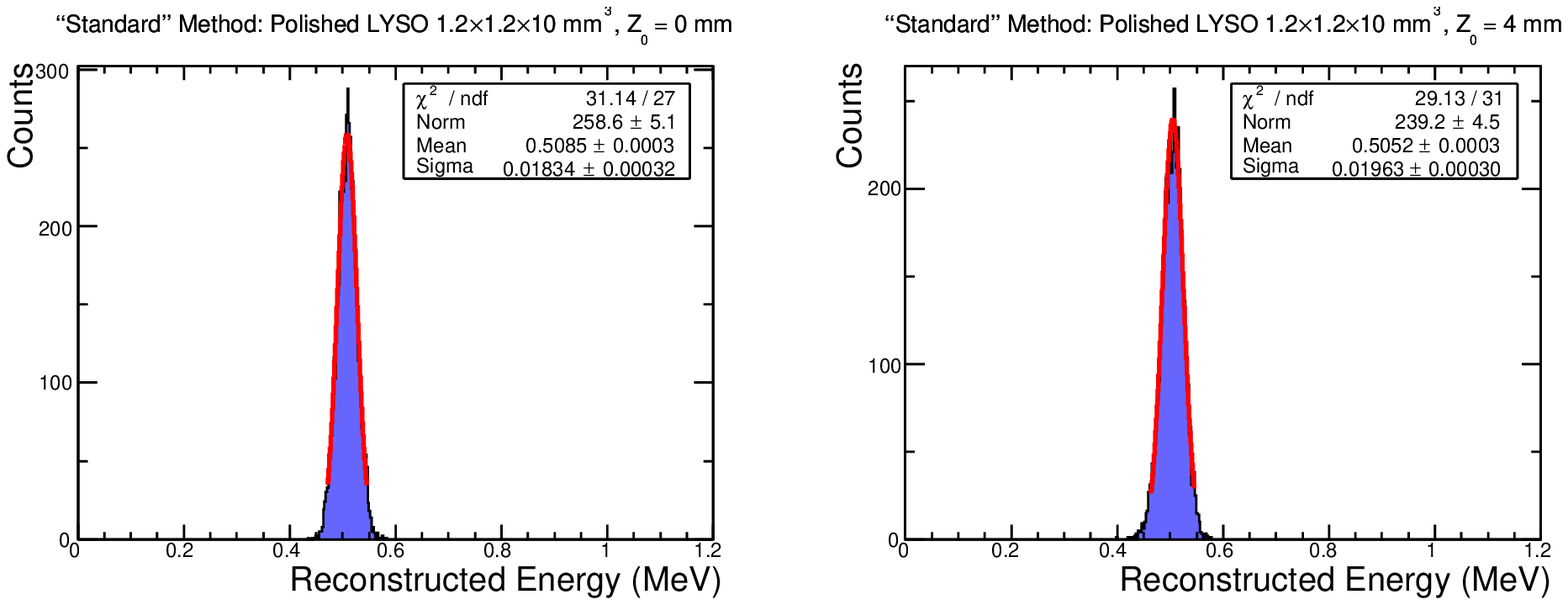}
 \end{center}
\caption{Distributions of energies that are reconstructed and calibrated with the ``standard'' ratio method for
1.2$\times$1.2$\times$10~mm$^3$ LYSO scintillator at $Z_0$ = 0 and 4~mm. }
\label{fig:standardenergy}
\end{figure}

The energy of the primary photon is reconstructed via a geometric mean of the amplitudes of the signals from both
edges of the LYSO scintillator:
\begin{equation}
E_{ NC} = \sqrt{Amp(+) \cdot Amp(-)}
\end{equation}
A mean value of the geometric-mean spectrum (normalized on the value
in the center of the scintillator) as a function of the primary photon position is shown in the left panel of Fig.~\ref{fig:standardenergyZ}. Because the
effective attenuation of the signal in the scintillator is not purely one-exponent effect, the residual dependence
of extracted but not-calibrated energy on the photon conversion position is observed; to ``calibrate out'' this
dependence, the left-panel plot in Fig.~\ref{fig:standardenergyZ} is fitted to the function:
\begin{equation}
e(Z) = p0 + p1 \cdot Z^2 + p2 \cdot Z^4 ,
\end{equation}
where $p0$, $p1$, and $p2$ are the fit parameters. Similar to the DOI calibration
(described in the beginning of this Section), I use the same simulation data for the calibration, but to calibrate
the real-life PET detector, the special calibration data taken with the collimated $\gamma$-source at few known
Z-positions will be needed. After the energy-calibration function $e(Z)$ is obtained, I use it to calibrate the
reconstructed energy on event-by-event basis:
\begin{equation}
Energy = (0.511 \ MeV) \cdot \sqrt{Amp(+) \cdot Amp(-)} / e(Z)
\end{equation}
A mean value of the calibrated-energy spectrum as a function of the primary photon position is shown in the right
panel of Fig.~\ref{fig:standardenergyZ}. With the described above energy-calibration method, the non-uniformity of the
calibrated energy over the scintillator length does not exceed 1\%.
Distributions of energies reconstructed and calibrated with the ``standard'' ratio method for
1.2$\times$1.2$\times$10~mm$^3$ LYSO scintillator at $Z_0$ = 0 and 4~mm are shown in
Fig.~\ref{fig:standardenergy}. One should understand that the simulated energy resolution of the detector reflects only the fluctuations in the photostatistics in the primary photon conversion and in the propagation and conversion  of ``optical'' photons; some signal fluctuations in the SiPMs and readout electronics are not included, so the energy resolution of the real-life detector might be a little worse (though I do believe that the difference should not be very big). 

\section{``New'' Method Results}

For the ``new'' method, I use the DOI and energy calibration procedures that are similar to the described before calibration procedures for the ``standard'' method.

\begin{figure}[htb]
\vspace*{58mm}
\begin{center}
\includegraphics{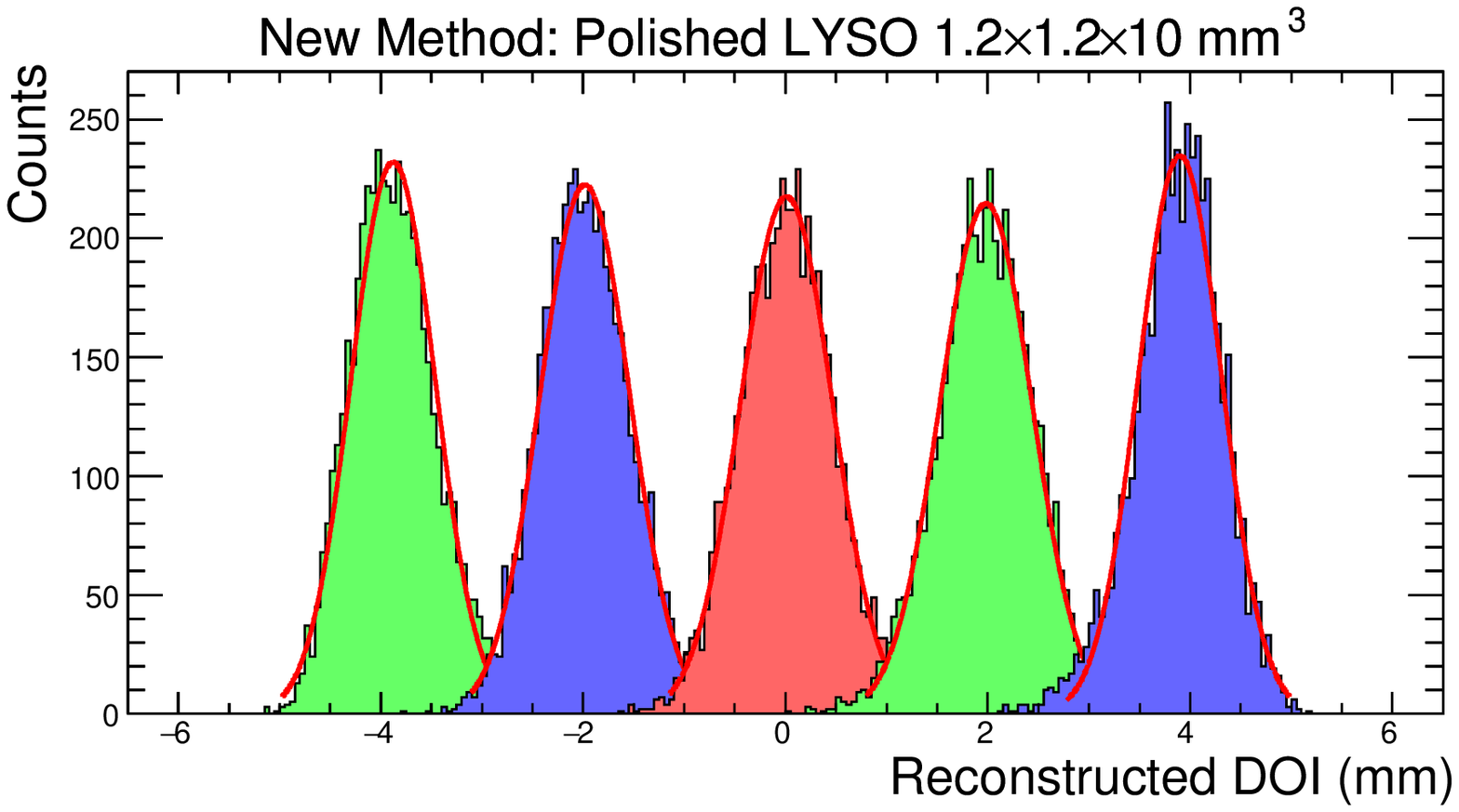}
 \end{center}
\caption{Depth-of-Interaction distributions that are reconstructed with the "new" method for
1.2$\times$1.2$\times$10~mm$^3$ LYSO scintillator at $Z_0$ = -4, -2, 0, 2, and 4~mm. 
(To be compared with Fig.~\ref{fig:standarddoi}.)}
\label{fig:doi10}
\end{figure}

DOI distributions that are reconstructed with the ``new'' method for 1.2$\times$1.2$\times$10~mm$^3$ LYSO
scintillator at $Z_0$ = 0 (center of the scintillator), $\pm$2~mm, and $\pm$4~mm (edges of the scintillator) 
are shown in Fig.~\ref{fig:doi10}. These ``new''-method distributions are more symmetric (viz., the asymmetric "tale" is greatly suppressed), and the widths of the distributions cores (DOI resolutions) are about 3-4 times better than the ones  from the ``standard'' ratio method. The DOI resolution ($\sigma$) that is achieved with the ``new'' method is better than 0.45~mm.

\begin{figure}[htb]
\vspace*{50mm}
\begin{center}
\includegraphics{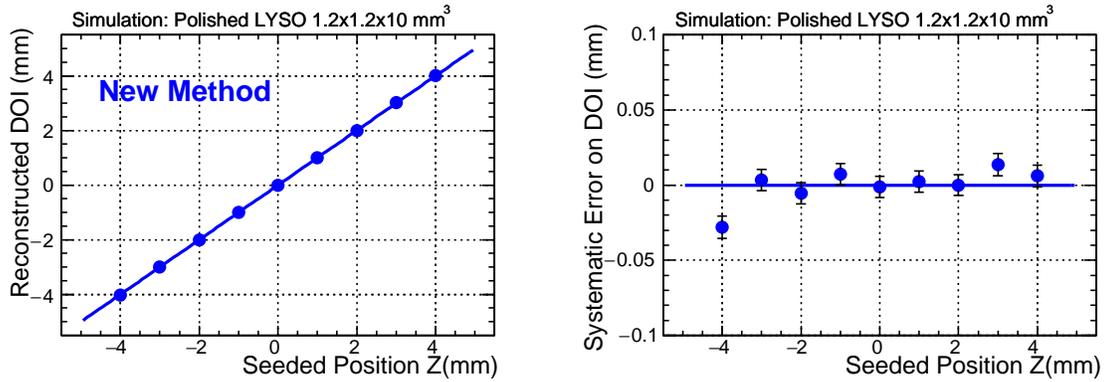}
 \end{center}
\caption{Left panel: The mean value for the Depth-of-Interaction distribution reconstructed with the "new" method
as a function of the seeded position Z (for 1.2$\times$1.2$\times$10~mm$^3$ LYSO scintillator). 
Right panel: Difference between the correspondent mean value for the DOI distribution reconstructed with the ``new''
method and the seeded position Z as a function of the seeded position Z.}
\label{fig:dZvsZ}
\end{figure}

The mean value for the DOI distribution reconstructed with the "new" method
as a function of the seeded position Z is shown in the left panel of the Fig.~\ref{fig:dZvsZ}. The right panel in the Figure shows a difference between the correspondent mean value for the DOI distribution reconstructed with the ``new'' method and the seeded position Z (viz., a systematic error that is introduces in the DOI reconstruction) as a function of the seeded position Z; as one can see from the Figure, this systematic error is consistent with zero.

\begin{figure}[htb]
\vspace*{55mm}
\begin{center}
\includegraphics{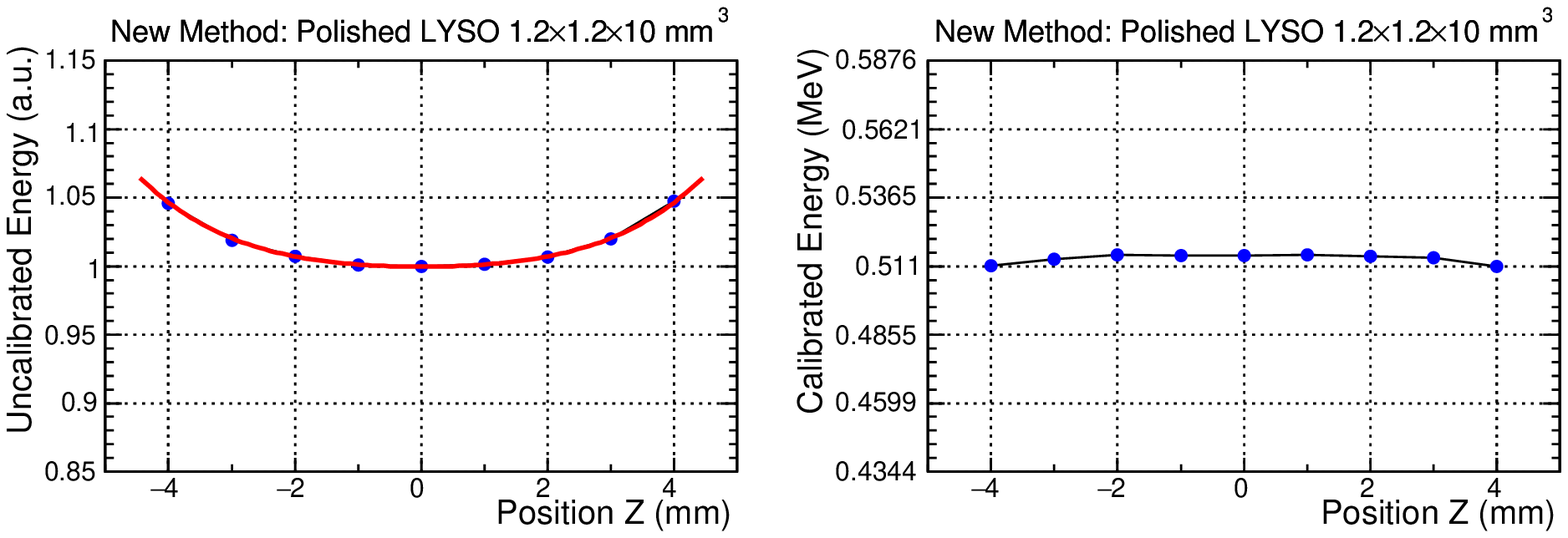}
 \end{center}
\caption{Average energy reconstructed with the ``new'' method before (left panel) and after calibration (right
panel) as a function of 0.511-MeV photon conversion position.}
\label{fig:standardenergyZ2}
\end{figure}

\begin{figure}[htb]
\vspace*{55mm}
\begin{center}
\includegraphics{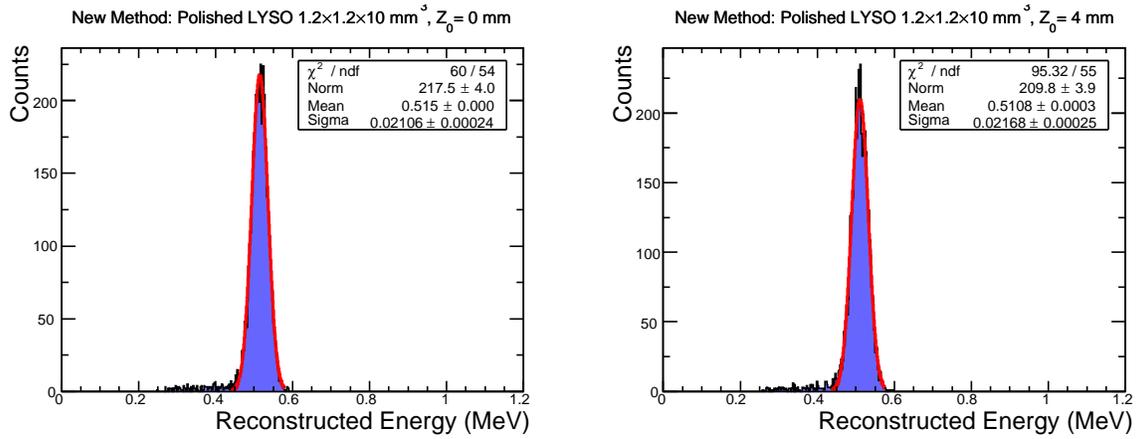}
 \end{center}
\caption{Distributions of energies reconstructed and calibrated with the ``new'' method for
1.2$\times$1.2$\times$10~mm$^3$ LYSO scintillator at $Z_0$ = 0 and 4~mm. }
\label{fig:standardenergy2}
\end{figure}

The mean values of the energy spectra before (left panel) and after the calibration (right panel) are shown in the Fig.~\ref{fig:standardenergyZ2}; the same as for ``standard'' ratio method, the non-uniformity of the
calibrated energy over the scintillator length does not exceed 1\%. The examples of the calibrated energy spectra at $Z_0$ = 0 and 4~mm are shown in Fig.~\ref{fig:standardenergy2}.

\begin{figure}[htb]
\vspace*{50mm}
\begin{center}
\includegraphics{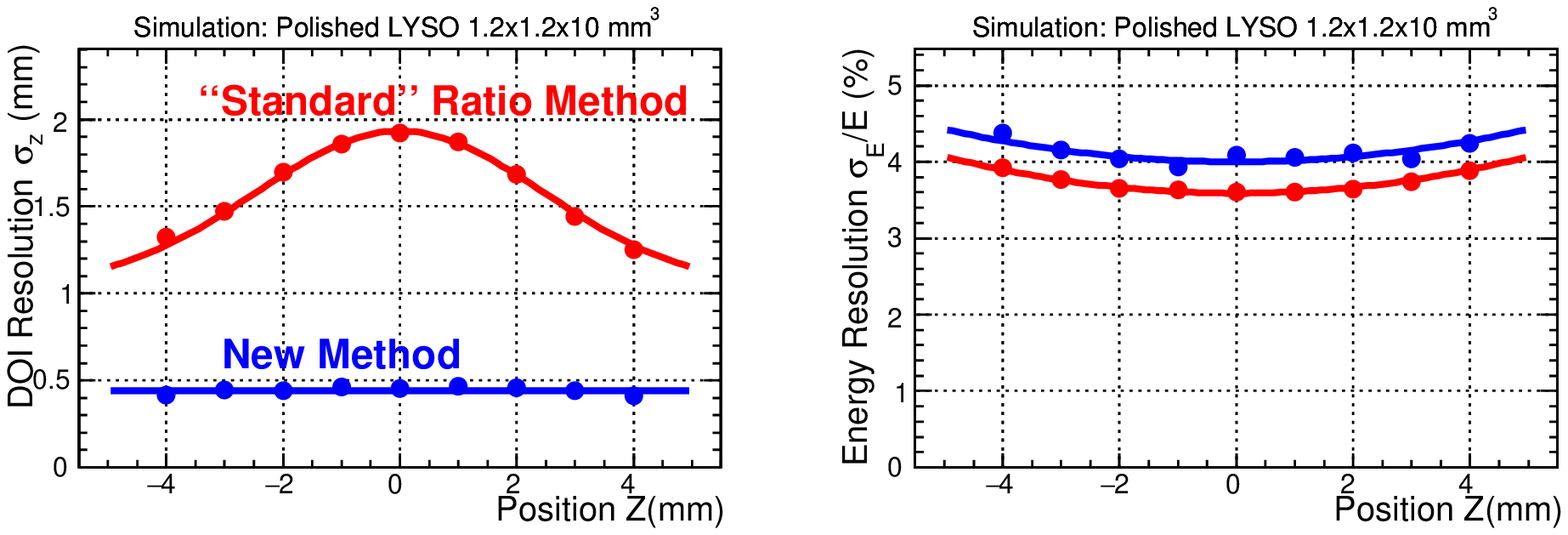}
 \end{center}
\caption{Left panel: Depth-of-Interaction resolutions (parameter $\sigma$ from the Gaussian fit) with 
the ``standard'' and the ``new'' methods for
1.2$\times$1.2$\times$10~mm$^3$ LYSO scintillator. Right panel: correspondent relative energy resolutions
($\sigma_E/E$).}
\label{fig:resol10}
\end{figure}

Fig.~\ref{fig:resol10} shows the summary comparison between the ``standard'' and ``new'' methods in terms of the DOI resolution (left panel) and the calibrated energy resolution (right panel). While the energy resolution for the ``new'' method is about 1.1 times worse than the one for the ``standard'' ratio method (that is not critical for PET), the ``new'' method has a big advantage (of about 3-4 times) in the DOI resolution. 

\section{``New DOI-Resolution-Enforced'' Method Results}

The described above ``new'' method allows significant improvement in the DOI resolution (compared to the ``standard'' ratio method) keeping the energy resolution almost the same good as for ``standard'' method. But if the precise measurement of the primary photon energy is not a high priority, the new method can be tuned to achieve even better results for the DOI resolution; I call this ``tuned'' method as the ``new DOI-resolution-enforced'' (NDRE) method. Again, the DOI and energy calibration procedures for the NDRE method are similar to the described before calibration procedures for the ``standard'' and ``new'' methods.

\begin{figure}[htb]
\vspace*{58mm}
\begin{center}
\includegraphics{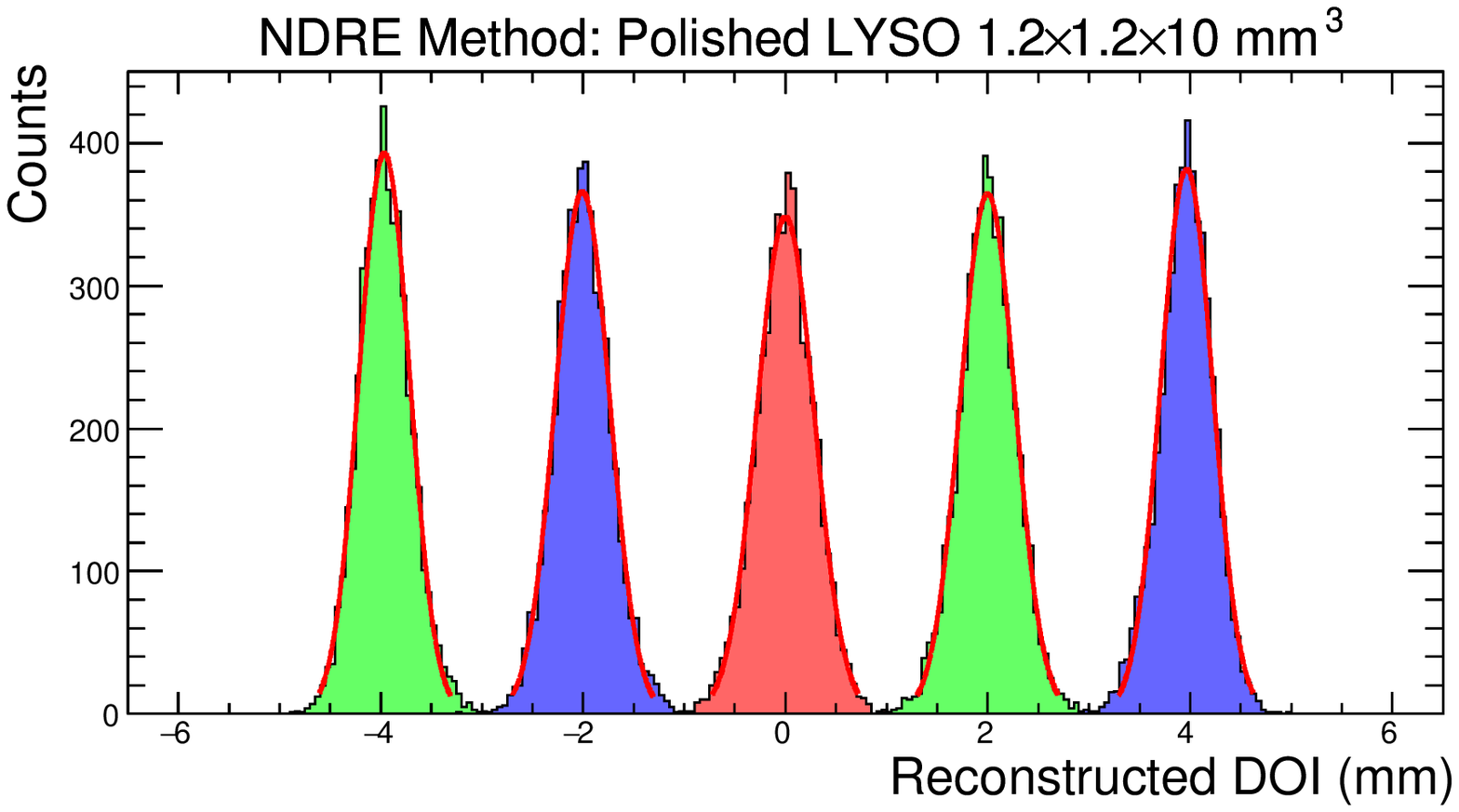}
 \end{center}
\caption{Depth-of-Interaction distributions reconstructed with the ``new 
DOI-resolution-enforced'' method for
1.2$\times$1.2$\times$10~mm$^3$ LYSO scintillator at $Z_0$ = -4, -2, 0, 2, and 4~mm. 
(To be compared with Fig.~\ref{fig:standarddoi} and Fig.~\ref{fig:doi10}.)}
\label{fig:doi10a}
\end{figure}

Fig.~\ref{fig:doi10a} shows DOI distributions that are reconstructed with the NDRE method for 1.2$\times$1.2$\times$10~mm$^3$ LYSO scintillator at $Z_0$ = 0 (center of the scintillator), $\pm$2~mm, and $\pm$4~mm. The DOI resolution ($\sigma$) that is achieved with the NDRE method is better than 0.3~mm that is about 1.5 times better compared to the ``new'' method.

\begin{figure}[htb]
\vspace*{50mm}
\begin{center}
\includegraphics{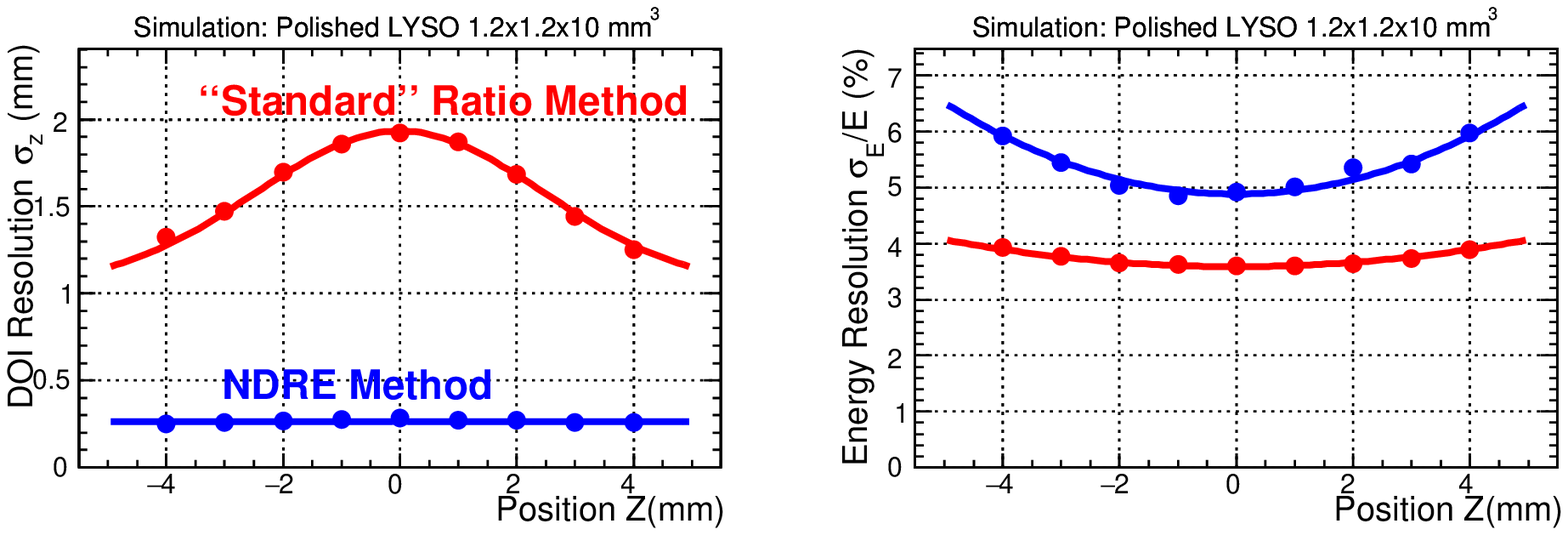}
 \end{center}
\caption{Left panel: Depth-of-Interaction resolutions (parameter $\sigma$ from the Gaussian fit) with the 
``standard'' and the ``new DOI-resolution-enforced'' methods for
1.2$\times$1.2$\times$10~mm$^3$ LYSO scintillator. Right panel: correspondent relative energy resolutions
($\sigma_E/E$). (To be compared with Fig.~\ref{fig:resol10}.)}
\label{fig:resol10a}
\end{figure}

The summary Fig.~\ref{fig:resol10a} shows the DOI and the calibrated energy resolutions for ``standard'' and NDRE methods as functions of the primary photon position. Compared to the ``standard'' method, the NDRE method provides about 5~(edge)-7~(center) times better DOI resolution and the calibrated energy resolution witch is worse by the factor of about 1.4-1.5~.  

\section{Resolutions with Longer LYSO Scintillators}

\begin{figure}[htb]
\vspace*{50mm}
\begin{center}
\includegraphics{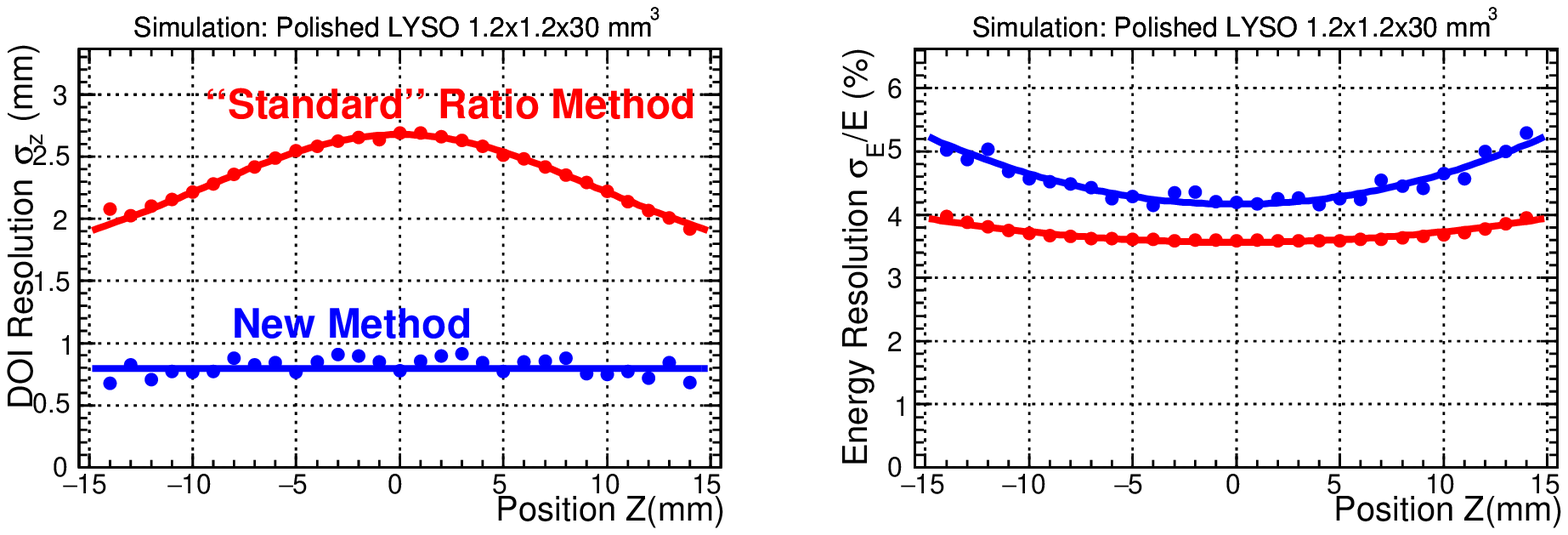}
 \end{center}
\caption{Left panel: Depth-of-Interaction resolutions (parameter $\sigma$ from the Gaussian fit) with the 
``standard'' and the ``new'' methods for
1.2$\times$1.2$\times$30~mm$^3$ LYSO scintillator. Right panel: correspondent relative energy resolutions
($\sigma_E/E$).}
\label{fig:resol30}
\end{figure}

\begin{figure}[htb]
\vspace*{50mm}
\begin{center}
\includegraphics{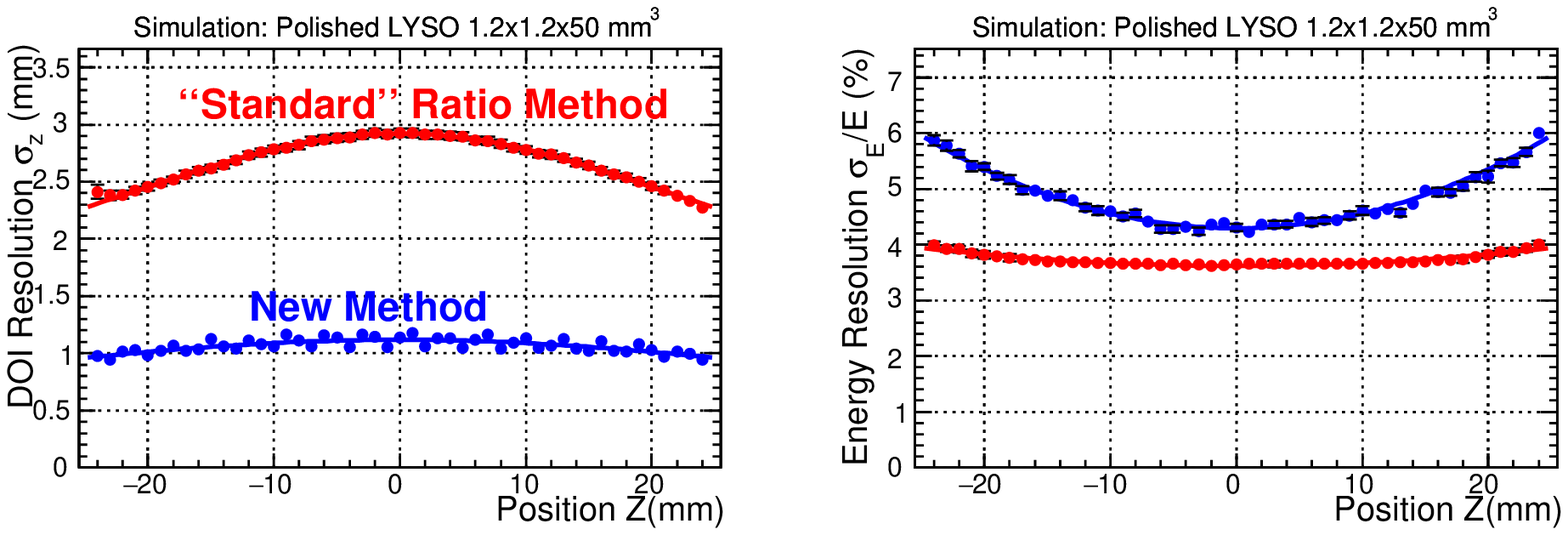}
 \end{center}
\caption{Left panel: Depth-of-Interaction resolutions (parameter $\sigma$ from the Gaussian fit) with the 
``standard'' and the ``new'' methods for
1.2$\times$1.2$\times$50~mm$^3$ LYSO scintillator. Right panel: correspondent relative energy resolutions
($\sigma_E/E$).}
\label{fig:resol50}
\end{figure}

Fig.~\ref{fig:resol30} shows the summary comparison between the ``standard'' and ``new'' methods in terms of the DOI resolution (left panel) and the calibrated energy resolution (right panel) for 1.2$\times$1.2$\times$30~mm$^3$ LYSO scintillator. Fig.~\ref{fig:resol50} shows the same summary for 1.2$\times$1.2$\times$50~mm$^3$ LYSO scintillator. 
Usage of the ``new'' method improves the DOI resolution by the factor of 2.7-3.0 for 30-mm-long scintillator, and by the factor of about 2.5 for 50-mm-long scintillator.

\section{Summary}

Table~1 contains the summary of DOI and energy resolutions.


\newpage

~\vspace*{2cm}

\vspace*{2cm}

~\\

\clearpage
\newpage

\begin{table}

\caption{DOI and relative energy resolutions for 1.2$\times$1.2$\times$Length LYSO scintillators. Only statistical uncertainties are shown.}
\begin{center}
\begin{tabular}{|l||c||c|c||c|c|} \hline
&&&&&\\
 Method/Position & Scintillator & DOI Resol. & DOI Resol. & Energy Resol. & 
 Energy Resol. \\
  & Length [mm] & ($\sigma$) [mm] & (FWHM) [mm] & ($\sigma$) [\%] & (FWHM) [\%] \\ 
  &&&&& \\ \hline
  &&&&&\\
 ``Standard''/edge & 10 & 1.30 $\pm$ 0.02 & 3.1 $\pm$ 0.1 & 3.6 $\pm$ 0.1 & 8.5 $\pm$ 0.3  \\ 
 ``Standard''/center & 10 & 1.92 $\pm$ 0.02 & 4.5 $\pm$ 0.1 &  3.6 $\pm$ 0.1 &  8.5 $\pm$ 0.3 \\ 
  &&&&&\\
 ``New''/edge & 10 & 0.41 $\pm$ 0.01 & 0.97 $\pm$ 0.03 & 4.1 $\pm$ 0.1 & 9.6 $\pm$ 0.3 \\
 ``New''/center & 10 & 0.45 $\pm$ 0.01 & 1.06 $\pm$ 0.03 & 4.1 $\pm$ 0.1 & 9.6 $\pm$ 0.3 \\
  &&&&&\\
 NDRE/edge & 10 & 0.25 $\pm$ 0.01 & 0.60 $\pm$ 0.02 & 5.4 $\pm$ 0.2 & 12.7 $\pm$ 0.4 \\
 NDRE/center & 10 & 0.28 $\pm$ 0.01 & 0.66 $\pm$ 0.02 & 4.9 $\pm$ 0.1 & 11.6 $\pm$ 0.3 \\
 &&&&&\\
 \hline
  &&&&&\\
 ``Standard''/edge & 30 & 1.95 $\pm$ 0.02 & 4.6 $\pm$ 0.1 & 3.6 $\pm$ 0.1 & 8.6 $\pm$ 0.3 \\
 ``Standard''/center & 30 & 2.70 $\pm$ 0.03 & 6.3 $\pm$ 0.1 & 3.6 $\pm$ 0.1 & 8.6 $\pm$ 0.3 \\
 &&&&&\\
 ``New''/edge & 30 & 0.72 $\pm$ 0.01 & 1.7 $\pm$ 0.1 & 4.2 $\pm$ 0.1 & 10.0 $\pm$ 0.3 \\
 ``New''/center & 30 & 0.90 $\pm$ 0.02 & 2.1 $\pm$ 0.1 & 4.1 $\pm$ 0.1 & 9.7 $\pm$ 0.3 \\
  &&&&&\\
 \hline
  &&&&&\\
 ``Standard''/edge & 50 & 2.35 $\pm$ 0.03 & 5.5 $\pm$ 0.1 & 3.7 $\pm$ 0.1 & 8.6 $\pm$ 0.3 \\
 ``Standard''/center & 50 & 2.93 $\pm$ 0.03 & 6.9 $\pm$ 0.1 & 3.6 $\pm$ 0.1 & 8.6 $\pm$ 0.3 \\
 &&&&&\\
 ``New''/edge & 50 & 0.95 $\pm$ 0.02 & 2.2 $\pm$ 0.1 & 4.8 $\pm$ 0.1 & 11.3 $\pm$ 0.3 \\
 ``New''/center & 50 & 1.14 $\pm$ 0.02 & 2.7 $\pm$ 0.1 & 4.3 $\pm$ 0.1 & 10.1 $\pm$ 0.3 \\
 &&&&&\\
 \hline
\end{tabular}
\end{center} 
\label{tab:1}
\end{table}

\end{document}